# Comet-like mineralogy of olivine crystals in an extrasolar proto-Kuiper belt

B. L. de Vries[1], B. Acke[1], J. A. D. L. Blommaert[1], C. Waelkens[1], L. B. F. M. Waters[2,3], B. Vandenbussche[1], M. Min[3], G. Olofsson[4], C. Dominik[3,5], L. Decin[1,3], M. J. Barlow[6], A. Brandeker[4], J. Di Francesco[7], A. M. Glauser[8,9], J. Greaves[10], P. M. Harvey[11], W. S. Holland[12,13], R. J. Ivison[12], R. Liseau[14], E. E. Pantin[15], G. L. Pilbratt[16], P. Royer[1] & B. Sibthorpe[12]

[1]Instituut voor Sterrenkunde, KU Leuven, Celestijnenlaan 200D, 3001 Leuven, Belgium.

[2]SRON, Sorbonnelaan 2, 3584 CA Utrecht, The Netherlands.

[3]Astronomical Institute "Anton Pannekoek", University of Amsterdam, PO Box 94249, 1090 GE Amsterdam, The Netherlands.

[4]Department of Astronomy, Stockholm University, AlbaNova University Center, 106 91 Stockholm, Sweden.

[5]Department of Astrophysics/IMAPP, Radboud University Nijmegen, PO Box 9010, 6500 GL Nijmegen, The Netherlands.

[6]Department of Physics and Astronomy, University College London, Gower St, London WC1E 6BT, UK.

[7]National Research Council of Canada, 5071 West Saanich Road, Victoria, British Columbia, V9E 2E7, Canada.

[8]ETH Zurich, Institute for Astronomy, Wolfgang-Paulistrasse 27, 8093 Zurich, Switzerland.

[9]UK Astronomy Technology Centre, Royal Observatory Edinburgh, Blackford Hill, Edinburgh EH9 3HJ, UK.

[10]SUPA, Physics and Astronomy, North Haugh, St Andrews, Fife KY16 9SS, UK.

[11]Astronomy Department, University of Texas, Austin, Texas 78712, USA.

[12]UK Astronomy Technology Centre, Royal Observatory, Blackford Hill, Edinburgh EH9 3HJ, Scotland, UK.

[13]Institute for Astronomy, University of Edinburgh, Royal Observatory, Blackford Hill, Edinburgh, EH9 3HJ, Scotland, UK.

[14]Earth and Space Sciences, Chalmers University of Technology, Onsala Space Observatory, 439 92 Onsala, Sweden.

[15]Laboratoire AIM, CEA/DSM-CNRS-Université Paris Diderot, IRFU/Service d'Astrophysique, Bâtiment 709, CEA-Saclay, 91191 Gif-sur-Yvette Cedex, France.

**Some planetary systems harbour debris disks containing planetesimals such as asteroids and comets[1]. Collisions between such bodies produce small dust particles[2], the spectral features of which reveal their composition and, hence, that of their parent bodies. A measurement of the composition of olivine crystals ($Mg_{2-2x}Fe_{2x}SiO_4$) has been done for the protoplanetary disk HD100546 (refs 3, 4) and for olivine crystals in the warm inner parts of planetary systems. The latter compares well with the iron-rich olivine in asteroids[5,6] ($x<0.29$). In the cold outskirts of the β Pictoris system, an analogue to the young Solar System, olivine crystals were detected[7] but their composition remained undetermined, leaving unknown how the composition of the bulk of Solar System cometary olivine grains compares with that of extrasolar comets[8,9]. Here we report the detection of the 69-micrometre-wavelength band of olivine crystals in the spectrum of β Pictoris. Because the disk is optically thin, we can associate the crystals with an extrasolar proto-Kuiper belt a distance of 15–45 astronomical units from the star (one astronomical unit is the Sun–Earth distance), determine their magnesium-rich composition ($x=0.01±0.001$) and show that they make up 3.6±1.0 per cent of the total dust mass. These values are strikingly similar to those for the dust emitted by the most primitive comets in the Solar System[8–10], even though β Pictoris is more massive and more luminous and has a different planetary system architecture.**

The olivine crystals found in the Itokawa asteroid and in ordinary chondrites (types 4 to 6) have an iron-rich composition[5] ($x < 0.29$). In contrast, laboratory measurements of olivine crystals from unequilibrated bodies such as comet 81P/Wild 2 and cometary interplanetary dust particles show that these crystals have a range of compositions, but the distribution has a pronounced and sharp peak at the almost pure magnesium-rich composition with $x<0.01$ (refs 8, 9). Both laboratory experiments[11] and observations[3,4] show that crystal formation in protoplanetary disks by gas-phase condensation, thermal annealing and shock heating results in magnesium-rich crystalline olivine[12–16] ($x<0.1$). During the protoplanetary disk phase, these olivine crystals are incorporated into planetesimals. An example of a planetary system in which the olivine crystals are then freed from such planetesimals by collisions is the system of β Pictoris. This system is a young (~12 Myr) analogue to the Solar System, with at least one planet at a distance of ~10 AU and a dusty debris disk containing small dust grains[7,17–20] (Fig. 1).

We have detected (Fig. 1) the 69 μm spectral band of small (~2 μm; see Supplementary Information), crystalline olivine grains in the planetary system of β Pictoris using Herschel[21] PACS[22]. The 69 μm band is of special interest because its exact peak wavelength and width are sensitive to both the grain temperature and, in particular, the composition of the olivine crystals[23,24] (Fig. 2). From our model fitting of the 69 μm band and spectral bands at shorter wavelengths (Fig. 1), the temperature and total mass of the crystals are determined to be 85±6 K and $(2.8±0.8) \times 10^{23}$ g, respectively. The exact wavelength position of the 69 μm band indicates very magnesium-rich crystalline olivine ($x=0.01±0.001$ (1σ)). The fraction of olivine crystals to the total amount of dust (obtained from the spectral energy distribution; see Supplementary Information) is 3.6±1.0% (1σ). The temperature of 85±6 K (1σ) places the population of olivine crystals between 15 and 45 AU from the central star, which coincides with a strong increase in surface density in the disk[25]. This location is outside the snow line of the system, where icy, comet-like bodies can exist, such as in the Kuiper belt of the Solar System. Scaling the distances in the β Pictoris system to those of the Solar System according to the different luminosities of the two central stars, the extrasolar Kuiper belt of β Pictoris reaches into the temperature range of the Jupiter–Saturn region. We propose that this location is an inward extension of what will in time become an analogue of the Kuiper belt of the Solar System.

The composition of the olivine crystals around β Pictoris is strikingly similar to that found in cometary bodies in the Solar System. From the low iron content, we can conclude that the olivine crystals we observe in β Pictoris come from collisions between unequilibrated, relatively small (<10 km) comet-like bodies[5]. The magnesium-rich olivine crystals around β Pictoris are in stark contrast to the iron-rich crystalline olivine[5] (x=0.29) found in asteroid-like bodies in the Solar System. When we compare the crystalline olivine abundance found in β Pictoris (3.6±1.0%) to that of primitive comets in the Solar System, similar low values are found. The comets 17P/Holmes and 73P/ Schwassmann–Wachmann, for example, contain about ~2–10% crys- talline olivine compared with the total amount of dust[10,26,27]. Because olivine crystals can be formed only within 10 AU of the central star[12–15] there must have been a transportation mechanism to bring these crystals to Kuiper belt distances. Studies of crystalline material and gas have indeed shown that radial mixing has taken place in both the Solar System and disks around young stars[28,29]. Models are able to predict crystalline olivine abundances of 2–58% at radii beyond 10AU on timescales of ~1Myr (refs 12, 30). The similar crystalline olivine abundances in β Pictoris and Solar System comets suggest that radial mixing must have been at work during the formation of the β Pictoris planetary system, with an efficiency similar to that in the protosolar nebula.

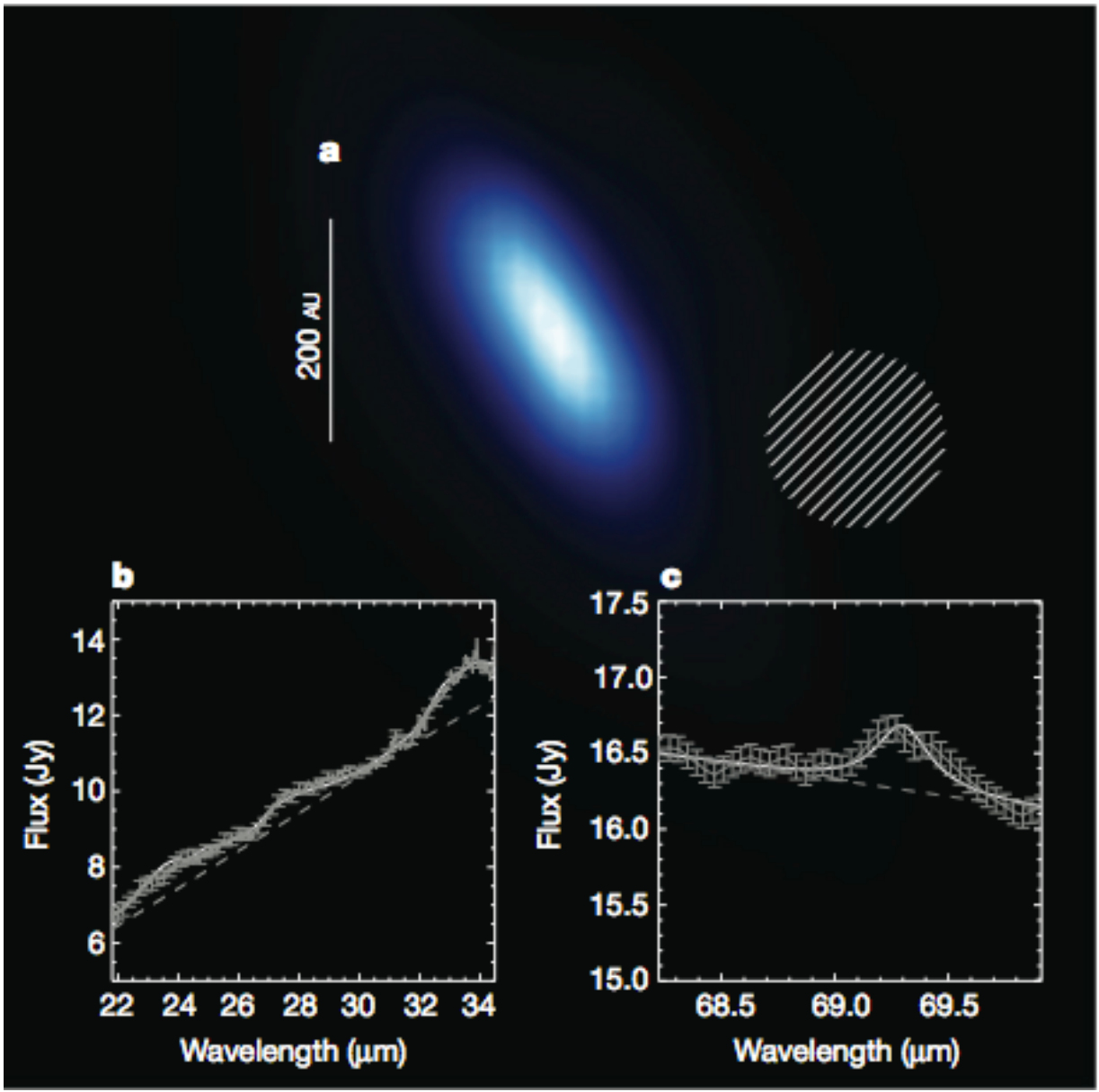

Figure 1 | Photometric and spectral observations of the planetary system of β Pictoris. a, Resolved surface brightness map of the β Pictoris debris disk at 70 mm taken with the Herschel Space Observatory's[21] Photodetector Array Camera and Spectrometer[22] (PACS). The disk is barely resolved with PACS, which has a point-spread function with a full-width at half-maximum of 8.299 (hatched circle). b, Spitzer Space Telescope infrared spectrograph spectrum showing prominent olivine features[7] (solid grey). The white solid line is our best model fit and the grey dashed line is the continuum. The uncertainties (1σ) in the Spitzer data[7] are indicated in the figure. c, The flux-corrected PACS spectrum with error bars (1σ) showing the 69 μm band of crystalline olivine (solid grey; 12σ detection). The white solid line shows the model fit to the 69 μm band of crystalline olivine as described in Supplementary Information, and the dashed grey line shows the underlying dust continuum. The best model contains crystalline olivine ($Mg_{2-2x}Fe_{2x}SiO_4$) with x=0.01±0.001 (1σ) and a temperature of 85±6 K (1σ).

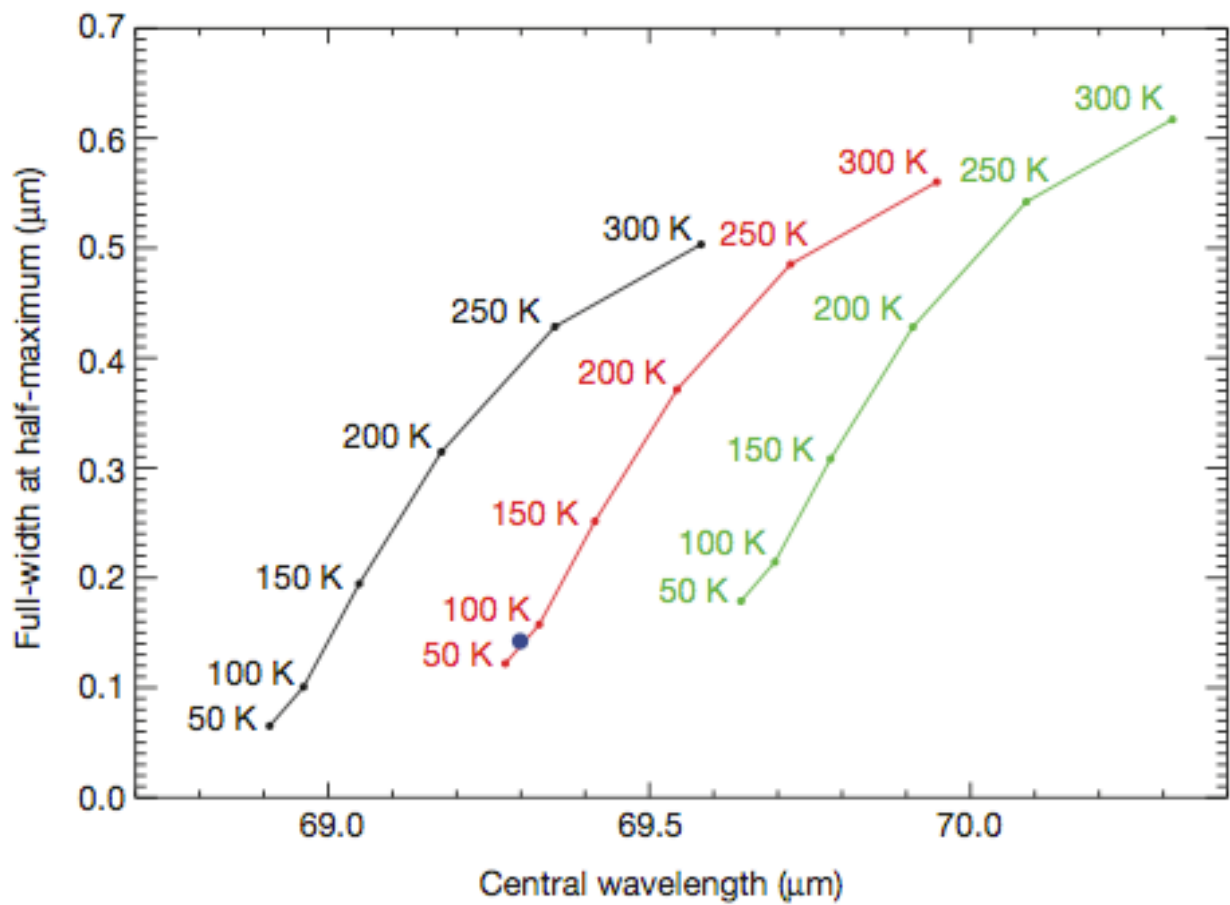

Figure 2 | Diagram demonstrating the dependence of the 69 μm band on grain temperature and composition. The diagram gives the width and central wavelength of the 69 μm band for six temperatures and for crystalline olivine ($Mg_{2-2x}Fe_{2x}SiO_4$) with x=0.0 (black), x=0.01 (red) and x=0.02 (green). The width and central wavelength are measured by fitting Lorentzian profiles to laboratory measurements[23,24] of crystalline olivine at different temperatures and compositions (see Supplementary Information for additional information). The width and wavelength positions measured show how the band broadens and shifts as a function of temperature or iron content. The best model fit of the 69 μm band of β Pictoris is indicated with a solid blue dot (Fig. 1b, c, white solid line); that is, the olivine crystals are cold (85±6 K) and contain about 1% iron (x=0.01±0.001).

**Acknowledgements**: Herschel is an ESA space observatory with science instruments provided by European-led Principal Investigator consortia and with important participation from NASA. PACS has been developed by a consortium of institutes led by MPE (Germany) and including UVIE (Austria); KUL, CSL and IMEC (Belgium); CEA and OAMP (France); MPIA (Germany); IFSI, OAP/AOT, OAA/CAISMI, LENS and SISSA (Italy); and IAC (Spain). This development has been supported by the funding agencies BMVIT (Austria), ESA-PRODEX (Belgium), CEA/CNES (France), DLR (Germany), ASI (Italy) and CICT/MCT (Spain). B.L.d.V. is an Aspirant Fellow of the Fund for Scientific Research, Flanders.

# Supplementary Information

## Observations

The *PACS* instrument[21] onboard *Herschel*[22] can be used as a photometer or spectrometer. Used as an integral field spectrometer it has a total field-of-view of 47" × 47", subdivided in a raster of 5 x 5 pixels of 9.4" × 9.4" each. On June 2, 2010 at 9:22:40 UTC (OD 384, OBSID 1342198172), we obtained a spectrum from 67.5 µm to 70.5 µm in the *PACS* Range Scan Mode. The integration time was 2.5 hours. The data have been reduced using the standard pipeline and *ipipe* scripts[30] (ChopNodRangeScan.py, v1.1 June 24, 2011). This reduction includes spectral flat-fielding to correct for small remaining scaling errors of the different spectral pixels.

Since the disk is barely resolved with the *PACS* spectrometer, no feature flux was seen in pixels surrounding the central one. We used the spectrum measured in the central 9.4" × 9.4" spatial pixel of the *PACS* integral field unit (See Fig.1 Main Article). The continuum level was further scaled to the 70 µm flux as measured in the *PACS* photometer map. The total flux at 70 µm with the *PACS* photometer is 16 Jy ± 0.8 Jy[20] (1σ).

The wavelength edges of the range scan are less well sampled, and therefore the signal-to-noise ratio is low. We discarded the part shortward of ~68.2 µm and longward of 70 µm. In addition, the infrared spectrum (~22 – 34 µm) obtained with the *Spitzer Space Telescope* was used in the model fitting[7]. Different slit measurements were combined to obtain a total area of 11" × 22", oriented along the disk. This area covers the entire disk.

## Optical constants and opacities

The interaction of light with dust particles is described by the opacity of the dust grains, which is the cross-section per mass unit ($cm^2/gr$). Laboratory studies either measure the absorbance of grains (which are transformed into opacities) or their refractive indices. In the latter case a particle model (including the grain size and shape) has to be adopted to calculate opacities from these indices. The opacities contain a contribution from scattering and absorption. For the micro-meter sized grains we are considering, the shape and strength of the 69 µm band are not sensitive to grain size, while those of crystalline olivine features in the 20-35 µm range are. From the shape of the 33.6 µm band, we deduced a grain size of ~2 µm. Figure S1 shows how the opacities of crystalline olivine around 33.6 µm change as a function of grain size. When the grain size increases, the 33.6 µm band develops a more pronounced red flank, making the feature broader. Fits to the spectrum show that the size of the crystalline olivine grains lies between 1 and 3 µm.

To determine the temperature and composition ($Mg_{(2-2x)}Fe_{2x}SiO_4$) of the crystalline olivine, 69 µm bands are simulated based on two dataset: a temperature dependent and a compositional dependent data set. The first data set[24] contains real and imaginary refractive indices of forsterite (x=0) at five different temperatures (50, 100, 150, 200, 295 K). The second set[23] contains absorption spectra of crystalline olivine at room temperature for different magnesium-iron compositions. Both sets include measurements of crystalline olivine at room temperature and with x=0. Comparing the sets at this point shows that both the width and the position of the 69 µm are the same within uncertainties. The measured strength of the two data sets is different, but this is probably due to the fact that the second set of measurements[23] are absorption measurements in KBr and polyethylene pellets while the first[24] are slap-reflection measurements.

To generate a 69 µm band at any temperature we fit the opacities of the 69 µm band generated from the temperature dependent set of measurements[24] with a Lorentzian. Opacities for the olivine crystals were calculated from the refractive indices using Gaussian Random Field[31] (GRF, porosity fraction of 0%) particle shapes. We then interpolate between the Lorentz parameters from the fit to obtain the wavelength position, width and strength of the band at any given temperature.

Simulating a 69 µm band with a different magnesium-iron composition than pure forsterite, the wavelength position, width and strength of the band are changed according to the given iron content. To know how much to change the Lorentz parameters of a pure forsterite band according to the iron content we want, we use the compositional dependent set of measurements[23]. We fit the 69 µm bands from this set[23] with Lorentzians and interpolate between the Lorentz parameters (see Fig S2 and S3). This gives us the information how much the Lorentz parameters change due to an increase in iron content.

Figure 2 (in the main article) summarizes how the width and position change when going from pure forsterite (x=0.0) to a magnesium-iron composition of x=0.01 or x=0.02, at different temperatures. From this figure, it is clear that even a small change of iron content of 1% induces a considerable and detectable shift in the position of the 69 µm band.

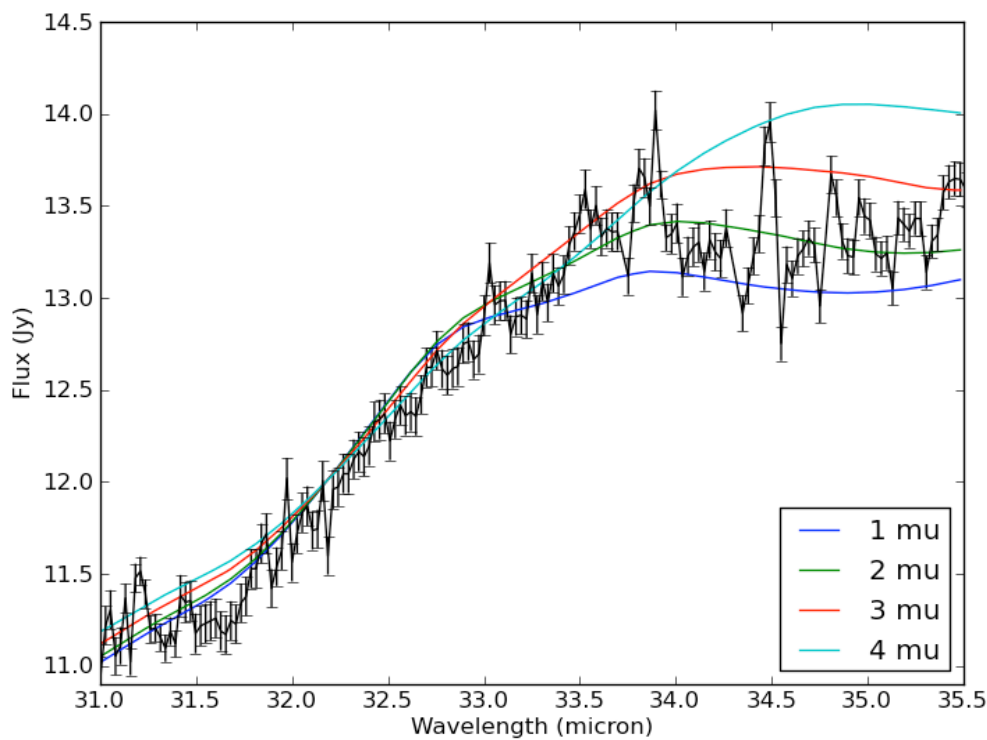

*Fig. S1: The dependence of the flux of crystalline olivine on grain size in the region of 33.6 µm.* *The red flank and peak of the feature are sensitive to the grain size such that the feature becomes bluer with decreasing grain size. Crystalline olivine grains with a radius between 1 and 3 µm provide the best fit to the observed 33.6 µm band shown in black[7] (1σ errors).*

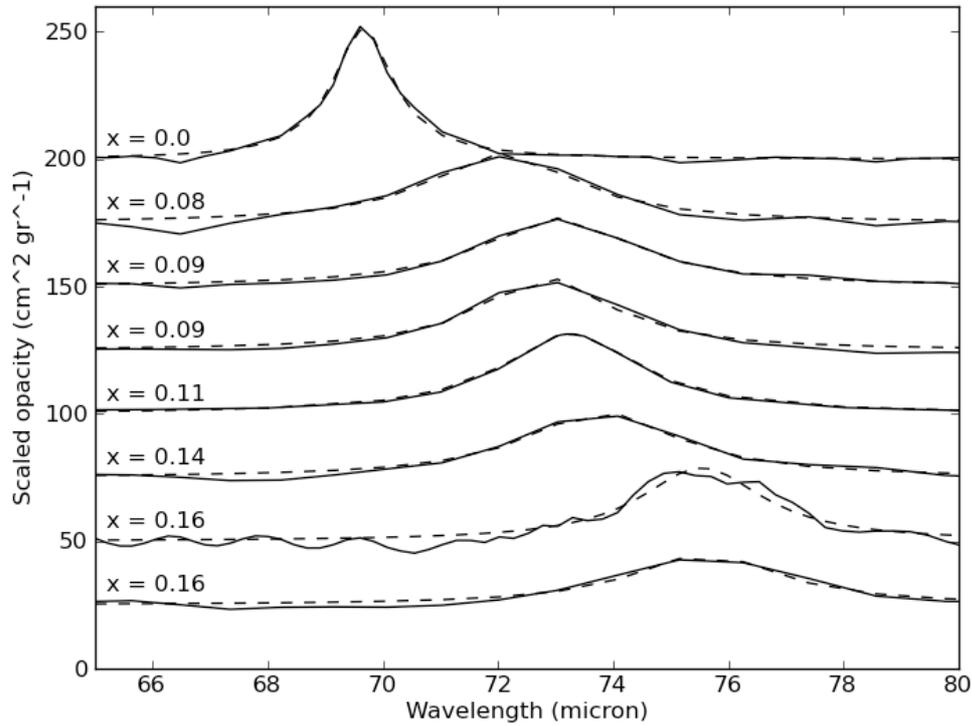

**Fig. S2: The opacity in the 69 μm bands for crystalline olivine with different magnesium-iron compositions**[23] **($Mg_{(2-2x)}Fe_{2x}SiO_4$).** The dashed lines show the Lorentzian fits to the 69 μm bands. The parameters of the fitted Lorentz curves are shown in Fig. S3.

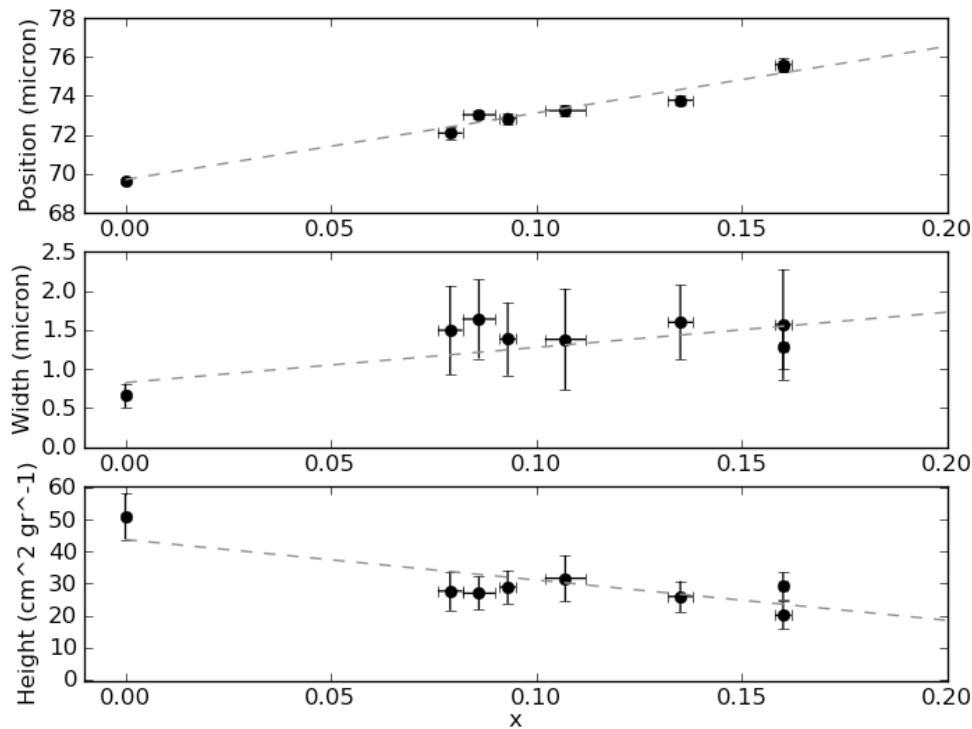

**Fig. S3: The dependence of the 69 μm band on the composition of the olivine crystals.** Panel a, b and c show how the position, width and the strength of the 69 μm band of crystalline olivine changes as a function of the magnesium-iron composition

($Mg_{(2-2x)}Fe_{2x}SiO_4$), respectively. These data (dots) have been generated by Lorentzian fits to laboratory data done at room temperature[23] (see Fig.S2). The errors on the three Lorentz parameters are obtained by a Monte Carlo simulation and adopting a 3% error (1σ) in the wavelength and opacity measurements[23]. Also included is the error in the iron content[23]. The dashed lines show linear fits to these data. The diagrams show that the position is the best and most precise probe of the magnesium-iron composition.

**Fit to the Spectral Energy Distribution of β Pictoris**
In this section we show the results from a fit to the spectral energy distribution (SED) at the wavelengths between 15 and 350 μm (see Fig. S4). We use *Spitzer* infrared data[7] and *PACS* photometry data[20]. The contribution of the central star to the SED has been subtracted from the spectrum. To fit the spectral energy distribution of the central star, an appropriate photosphere model[32] was scaled to literature multi-color photometry. The photometry covers a wavelength range from the ultraviolet region to the near-infrared.

For the SED fit we use a three temperature optically thin model. Here the optical thickness τ of the medium is defined as the integral of $d\tau = \rho \cdot \kappa \cdot dr$, where dr walks over the line of sight of interest. The optically thin case is where τ<<1. For the opacities we use dust particles with a size ranging from 2 to 100 μm and a size distribution power-law with an index of -3.5. The dust particles used in the SED fit are taken as a porous (porosity fraction 50%) composite grain with a typical interplanetary dust particle (IDP) composition[33], containing 45% ice, 10% FeS, 32% amorphous $MgFeSiO_4$ and 13% carbon. A $\chi^2$ minimalization fit to the SED, using the same methodology as described in the section "Fitting of the crystalline olivine features", results in a small (($5.5 \pm 2.7$) × $10^{21}$ g) but hot population (436 K ± 85 K) of grains, which must be close to the central star, inside the depleted region (< 30 AU). A population three orders of magnitude larger (($7.9 \pm 0.3$) × $10^{24}$ g) is found at 94 K ± 2 K. And a third population of ($1.8 \pm 0.1$) × $10^{26}$ g is found at 22 K ± 1 K.

The errors (1σ) in the masses and temperatures are obtained by repeating the fit multiple times (MC simulation, 100 repetitions), while randomly varying the composition of the composite grains. The composition of the grains can vary by 5% (1σ) at each iteration.

Note that the flux in the 20-100 μm region is dominated by the 94 K dust population, which shows that the flux in this region can be described by one temperature.

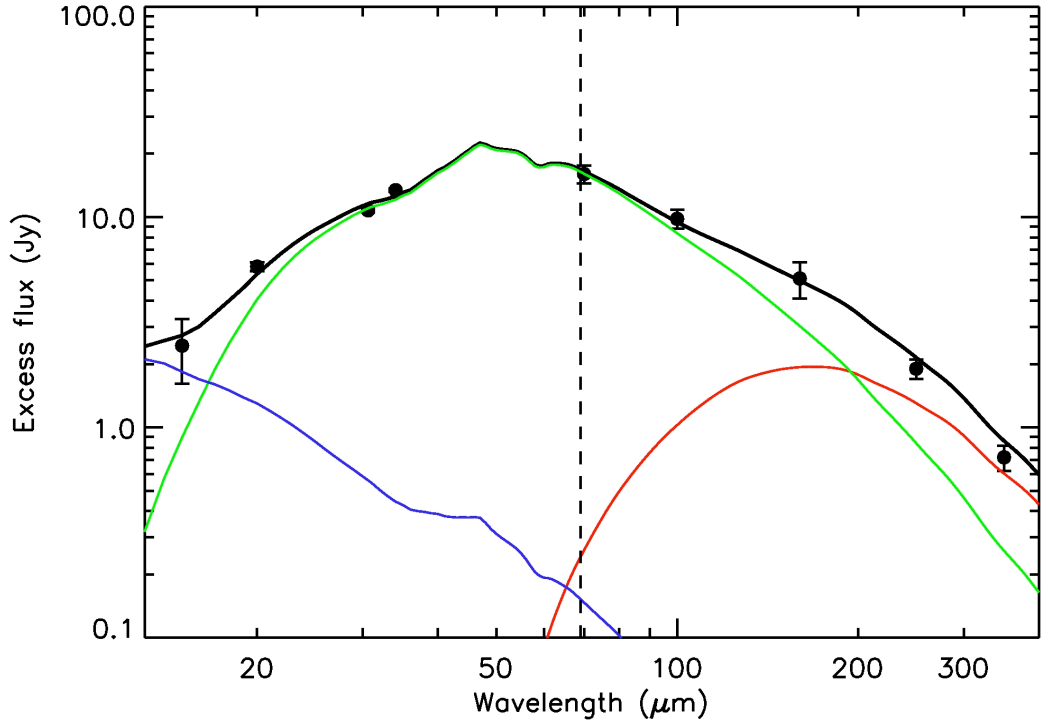

**Fig. S4: Fit to the SED of β Pictoris.** The fluxes as measured with *Spitzer*[7] and the *PACS* Photometer[20] are shown as dots, the best model fit as a black line and the contributions from the three different temperatures: 436 K in blue, 94 K in green and 22 K in red. Shown are the error bars (1σ) for the *Spitzer* data[7] and *PACS* photometry[20].

**Fitting of the crystalline olivine features**

We have fitted the spectral features in the 20-34 µm range and the 69 µm band simultaneously. This approach enables us to constrain uniquely the temperature, magnesium-iron composition and total mass of the crystalline olivine material. The dust in the disk is optically thin at all wavelengths. To show this, we determine an upper limit to the optical depth at 69 µm ($\tau_{69}$) in the radial direction (β Pictoris is seen edge-on[34]). A maximal optical depth at 10 µm vertically through the disk (parallel to the disk axes) at a distance from the star of ~80 AU is found[34] to be $\tau'_{10} = 10^{-2}$. Taking an opening angle for the disk of 30 degrees[34], assuming a constant density in the disk and taking the disk size to go from 30-200 AU[20], the optical depth in the radial direction through the disk has an upper limit of $\tau_{10} = 4 \times 10^{-2}$. Using the composite particles also used in the SED fit in the section "Fit to the Spectral Energy Distribution of β Pictoris", the opacities at 69 µm are a factor ~5 lower than those at 10 µm, leading to an upper limit of the optical depth of $\tau_{69} \sim 8 \times 10^{-3}$.

Since the disk is optically thin, we do not depend on a model of the specific disk geometry, but can simply describe the emission of the olivine grains by $B_\nu(T) \cdot M \cdot \kappa_\nu(T,x)$, with $B_\nu(T)$ the Planck function at a grain temperature T, M the total emitting mass times the solid angle of the emitting surface and $\kappa_\nu(T,x)$ the opacities of olivine crystals at grain temperature T and with a composition of $Mg_{(2-2x)}Fe_{2x}SiO_4$. A one temperature fit in this wavelength range is warranted (see section "Fit to the Spectral Energy Distribution of β Pictoris"). The opacities $\kappa_\nu(T,x)$ are generated using GRF particles[31] at a size of 2 µm (porosity factor 0%, see Fig.S1). The parameters M, T and x that determine the spectral features are fitted simultaneously with the underlying dust continuum. To obtain the best fit, we start with an evolutionary algorithm to find an optimal location in the multiple parameter space. This evolutionary algorithm is designed to check solutions against local minima. Subsequently, the parameters returned by the evolutionary algorithm are used as starting parameters for a $\chi^2$ minimization algorithm to fine-tune the fit.

Confidence intervals on the best-fit parameters were derived via a Monte Carlo simulation. We repeated the fitting to the spectra 2500 times, and at every iteration different uncertainties (see below) were randomly added to the data. This technique yields distributions for the fitting parameters. The standard deviations are used as a measure for the uncertainty on the derived parameter. The following uncertainty sources were taken into account:

- The uncertainty on the data points of the Spitzer spectrum[7].
- The relative uncertainty in each wavelength pixel in the PACS range-scan is estimated at every MC iteration. We fitted a line through the continuum part of the PACS scan in the 67.7 - 68.8 µm range. The noise is estimated by taking the standard deviation around this linear fit. The average noise is 0.13 Jy per spectral pixel (1σ).
- A 10% uncertainty on the absolute flux calibration of the PACS range scan and the uncertainty in the PACS Photometry at 70 µm (0.8 Jy, 1σ)[20].

The parameters and uncertainties (1σ) of the best fit to the spectral features are a grain temperature of 85 ± 6 K, a composition of x=0.01 ± 0.001, and a mass of $(2.8 ± 0.8) \times 10^{23}$ g.

The significance of the 69 µm band is estimated during the MC simulations. For each simulation, the integrated flux of the band is measured. From the distribution of this quantity, we obtain 0.16 ± 0.013 Jy×micrometer, i.e. the 69 µm feature is a 12 σ detection.

If we take the total mass of the crystalline olivine population obtained from the fit to the spectral features of crystalline olivine ($(2.8 ± 0.8) \times 10^{23}$ g) and compare that to the total dust mass of the 94 K population obtained from the SED fit ($(7.86 ± 0.26) \times 10^{24}$ g), we find an abundance of the crystalline olivine compared to the total dust mass of 3.6 % ± 1.0 % (1σ). This means that, taking an IDP composition for the dust, a fraction of 10 % ± 3 % of the olivine material is in crystalline form.

With these results we repeated the SED fit in the section "Fit to the Spectral Energy Distribution of β Pictoris", now including 3.6 % olivine crystals. This alters the composition for the dust grains to 43.4 % ice, 9.6 % FeS, 30.9 % amorphous $MgFeSiO_4$ and 12.5 % carbon, and 3.6 % crystalline olivine. This did not result in a significant difference for the best-fit parameters in the SED fit, however.

The iron content we find and the fact that the crystalline olivine is so magnesium rich can now be compared to what is found in cometary material in our Solar System[8,9,26,35-37].

**Geometrical location of the olivine crystals**

Using the derived temperature of the crystalline olivine grains (85 K ± 6 K), the distance of the small dust grains to the central star can be estimated. We calculate the temperature of the dust in the disk using the radiative transfer code MCMax[38]. To fit the spectral energy distribution of the central star, an appropriate photosphere model[32] was scaled to literature multi-color photometry, covering a wavelength range from the ultraviolet region to the near-infrared.

From the fit to the features we find that the crystalline olivine has almost the same temperature as the grains producing the underlying dust continuum, obtained from the SED fit. From this we find that the most likely situation is that the crystalline olivine grains are incorporated into the larger grains causing the dust continuum, meaning that the temperature of all minerals is more or less equal. Therefore, to derive the dust temperature in the disk, we consider a composite grain in which the small crystalline olivine grains are embedded[33]. Since micron-sized crystalline olivine grains are almost transparent at the wavelengths at which the central star emits most of its energy, the other species contained in the grain dominate the heating of the grain. We used opacities of grains with a typical interplanetary dust particle composition[33], which contain 43.4 % ice, 9.6 % FeS, 30.9 % amorphous MgFeSiO$_4$ and 12.5 % carbon and 3.6 % crystalline olivine.

Two extreme cases for the composite grain size were considered; at the small end, we use 2 µm, which is the estimated size of the olivine crystals themselves and hence a lower limit to the size of the composite grains. We also compute the temperature distribution in the disk assuming 100-µm-sized grains. Figure S5 shows the dust temperature in the disk of β Pictoris using these grain sizes. The figure shows that the position of the olivine crystals in the disk is between 15 and 45 AU. This distance corresponds to the location where the surface density increases outwards in the disk[25,34].

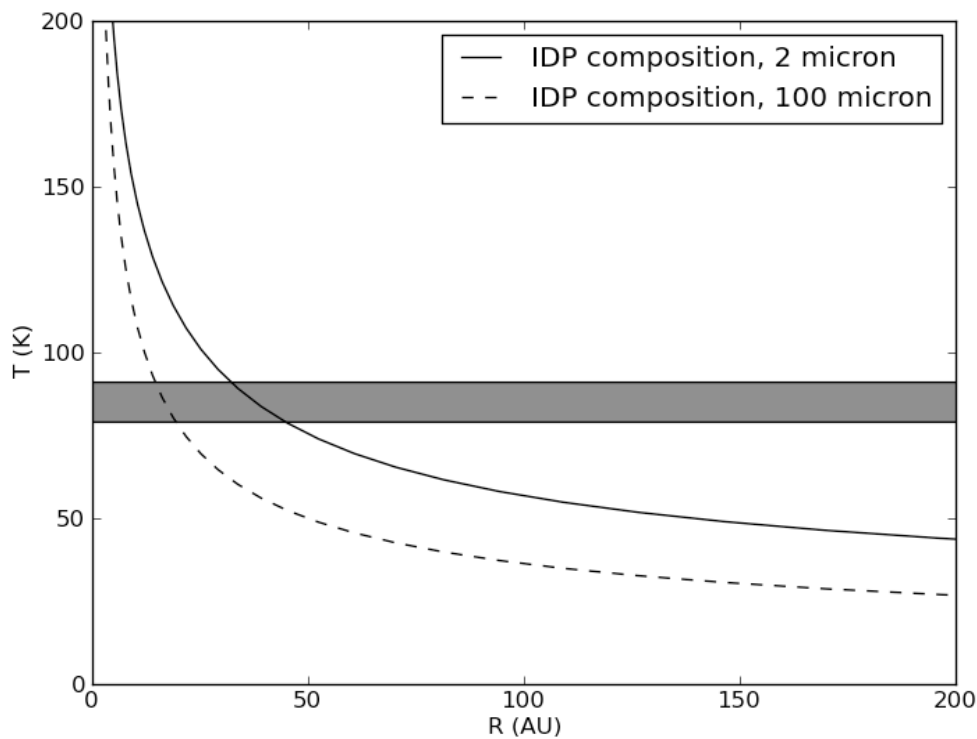

**Fig. S5: Plot demonstrating how the temperature of a grain with typical IDP composition varies as function of distance to the central star.** We assume, based on the results from the SED fit, that the crystalline olivine is part of such an IDP grain. To calculate the temperature structure in the disk, we used a stellar model with an age of 12 Myr and 9.2 L$_{solar}$[39,40]. The temperature structure has been calculated for two particle sizes: 2 µm (solid line) and 100 µm (dotted line). The horizontal grey box indicates the temperature obtained from the best fit, ranging from 79 K to 91 K. This shows that the observed olivine crystals must be located at ~15-45 AU from the central star.